\documentclass{aa}  

\usepackage{graphicx}
\usepackage{txfonts}
\begin{document}

   \title{Optical characterization of WISE selected blazar candidates}
   \titlerunning{Characterization of WISE blazar candidates}

   \author{Raniere de Menezes$^{1,2}$\thanks{E-mail: raniere.m.menezes@gmail.com}, Harold A. Pe\~na-Herazo$^{1,3,4,5}$, Ezequiel J. Marchesini$^{1,4,7,8,9}$, Raffaele D'Abrusco$^6$, Nicola Masetti$^{9,10}$, Rodrigo Nemmen$^2$, Francesco Massaro$^{1,4,5,13}$,  Federica Ricci$^{11}$, Marco Landoni$^{12}$, Alessandro Paggi$^{4,5}$, and Howard A. Smith$^6$
          }
   \authorrunning{R. de Menezes et al.}

   \institute{Dipartimento di Fisica, Universit\`a degli Studi di Torino, via Pietro Giuria 1, I-10125 Torino, Italy
   \and
   Universidade de S\~ao Paulo, Departamento de Astronomia, S\~ao Paulo, SP 05508-090, Brazil
   \and
   Instituto Nacional de Astrof\'isica, \'Optica y Electr\'onica (INAOE) Apartado Postal 51-216, 72000 Puebla, Mexico
   \and
   INFN -- Istituto Nazionale di Fisica Nucleare, Sezione di Torino, via Pietro Giuria 1, I-10125 Turin, Italy
   \and
   INAF-Osservatorio Astrofisico di Torino, via Osservatorio 20, 10025 Pino Torinese, Italy
   \and
   Center for Astrophysics | Harvard \& Smithsonian, 60 Garden Street, Cambridge, MA 20138, USA
   \and
   Facultad de Ciencias Astron\'omicas y Geof\'isicas, Universidad Nacional de La Plata, Paseo del Bosque, B1900FWA, La Plata, Argentina
   \and
   Instituto de Astrof\'isica de La Plata, CONICET--UNLP, CCT La Plata, Paseo del Bosque, B1900FWA, La Plata, Argentina
   \and
   INAF -- Osservatorio di Astrofisica e Scienza dello Spazio, via Gobetti 93/3, I-40129, Bologna, Italy
   \and
   Departamento de Ciencias F\'isicas, Universidad Andr\'es Bello, Fern\'andez Concha 700, Las Condes, Santiago, Chile
   \and
   Instituto de Astrof\'isica and Centro de Astroingenier\'ia, Facultad de F\'isica, Pontificia Universidad Cat\'olica de Chile, Casilla 306, Santiago 22, Chile
   \and 
   INAF - Osservatorio Astronomico di Brera, via E Bianchi 46 Merate (LC) Italy
   \and 
   Consorzio Interuniversitario per la Fisica Spaziale (CIFS), via Pietro Giuria 1, 10125 Torino, Italy
             }

   \date{Received June 27, 2019; accepted August 14, 2019}

 
  \abstract
   {Over the last decade more than five thousand $\gamma$-ray sources were detected by the Large Area Telescope (LAT) on board \textit{Fermi} Gamma-ray Space Telescope. Given the positional uncertainty of the telescope, nearly $30\%$ of these sources remain without an obvious counterpart in lower energies. This motivated the release of new catalogs of $\gamma$-ray counterpart candidates and several follow up campaigns in the last decade.}
   {Recently, two new catalogs of blazar candidates were released, they are the improved and expanded version of the WISE Blazar-Like Radio-Loud Sources (WIBRaLS2) catalog and the Kernel Density Estimation selected candidate BL Lacs (KDEBLLACS) catalog, both selecting blazar-like sources based on their infrared colors from the Wide-field Infrared Survey Explorer (WISE). In this work we characterized these two catalogs, clarifying the true nature of their sources based on their optical spectra from SDSS data release 15, thus testing how efficient they are in selecting true blazars.}
   {We first selected all WIBRaLS2 and KDEBLLACS sources with available optical spectra in the footprint of Sloan Digital Sky Survey data release 15. Then we analyzed these spectra to verify the nature of each selected candidate and see which fraction of the catalogs is composed by spectroscopically confirmed blazars. Finally, we evaluated the impact of selection effects, specially those related to optical colors of WIBRaLS2/KDEBLLACS sources and their optical magnitude distributions.}
   {We found that at least $\sim 30\%$ of each catalog is composed by confirmed blazars, with quasars being the major contaminants in the case of WIBRaLS2 ($\approx 58\%$) and normal galaxies in the case of KDEBLLACS ($\approx 38.2\%$). The spectral analysis also allowed us to identify the nature of 11 blazar candidates of uncertain type (BCUs) from the \textit{Fermi}-LAT 4th Point Source Catalog (4FGL) and to find 25 new BL Lac objects.}
   {}


   \maketitle
%

\section{Introduction}
\label{Intro}

One of the main challenges of modern $\gamma$-ray astronomy in the era of \textit{Fermi} Large Area Telescope (LAT) is the association of $\gamma$-ray sources with their low-energy counterparts \citep{massaro2015refining,massaro2016extragalactic}. The main problem underlying it is the large positional uncertainty of $\gamma$-ray detected sources, typically $\sim 4'$ in the \textit{Fermi}-LAT 4th Point Source Catalog\footnote{\url{https://fermi.gsfc.nasa.gov/ssc/data/access/lat/8yr_catalog/}} \citep[4FGL;][]{lat20194fgl}. Association task also affects source classification that covers up to $\sim 70\%$ of the 4FGL leaving 1521 unidentified/unassociated $\gamma$-ray sources (UGSs) to date.

UGSs are distributed throughout the whole sky uniformly -- although showing some concentration towards the Galactic plane ($|b| < 20^{\circ}$) -- indicating that most of them could have extragalactic nature. The $\gamma$-ray sky is mainly populated by non-thermal sources and in particular $\sim 80\%$ of 4FGL associated sources are classified as blazars. These belong to one of the rarest class of active galactic nuclei (AGNs), whose emission arises from particles accelerated in a relativistic jet closely aligned with line of sight \citep{blandford1978pittsburgh}. Thus it is expected that a significant fraction of UGSs is composed of blazars, at least at high Galactic latitudes.

Blazars are divided into BL Lacs (BZBs) and blazars of quasar type, labelled in the Roma-BZCAT as BZQs \citep{massaro2015romabzcat5th}, and classified based on their optical spectra, where the former have featureless optical spectra, or only absorption lines of galactic origin and weak and narrow \citep[$< 5$\AA;][]{stickel1991complete,landoni2014spectroscopy,landoni2015redshift,landoni2015optical} emission lines; and the latter have flat radio spectra, with optical spectra showing broad emission lines and a dominant intrinsically blue continuum. Blazars, emitting non-thermal radiation over the whole electromagnetic spectrum, also show large variability at all wavelengths, a flat or inverted radio spectrum, significant polarized emission and in some cases even apparent super-luminal motion \citep{urry1995unified,abdo2010blazars}. 

Blazars occupy a specific region in the mid-infrared (IR) color space defined by the Wide-field Infrared Survey Explorer filters~\citep[WISE;][]{wright2010WISE_telescope}. This region is known as the WISE Blazar Strip \citep{massaro2011identification,massaro2012wise,dAbrusco2012infrared} when it is described
in a two-dimensional space, and {\it locus} when it is modeled in the full three-dimensional WISE color space~\citep{dAbrusco2014wibrals}. Such distinctive mid-IR colors are due to the non-thermal emission arising from their relativistic jets \citep{bottcher2007modeling}. This discovery led to search for blazar-like sources lying within the positional uncertainty ellipses of UGSs that could be their potential counterparts \citep{dAbrusco2014wibrals,massaro2015refining}.

One of the catalogs that provides the largest number of associated blazar candidates for \emph{Fermi} catalogs is the WISE Blazar-Like Radio-Loud Sources catalog \citep[WIBRaLS;][]{dAbrusco2014wibrals,dAbrusco2019wibrals2_KDEBLLACS}. Selected sources in the WIBRaLS are detected in all four mid-IR WISE bands (nominally at 3.4, 4.6, 12 and 22 $\mu$m) and show mid-IR colors similar to confirmed \textit{Fermi} blazars. Sources in WIBRaLS are also required to have a radio-loud counterpart (see more details in Section \ref{Sample:WIBRaLS}). In the latest release of the WIBRaLS catalog (hereinafter WIBRaLS2), an additional list of candidate BZBs has been built based on a kernel density estimation (KDE) technique, namely KDEBLLACS \citep{dAbrusco2019wibrals2_KDEBLLACS}. Sources in KDEBLLACS are also radio-loud. Sources in both catalogs are selected with Galactic latitude $|b| > 10^{\circ}$.

The main goal of the present analysis is to characterize sources listed in both WIBRaLS2 and KDEBLLACS, on the basis of the optical spectra available for those lying in the footprint of the Sloan Digital Sky Survey \citep[SDSS;][]{york2000sloan} 15th data release \citep[DR15;][]{aguado2019fifteenth}. We aim to classify them spectroscopically mainly to determine the source classes that contaminate the WIBRaLS2 selection criteria. SDSS is ideal to achieve our purposes since it is the survey with higher number of spectroscopic objects available in the literature.

The paper is organized as follows: in Section \ref{SampleSelection} we describe samples used to carry out our analysis, providing basic details of both WIBRaLS2 and KDEBLLACS, selecting those sources in the SDSS footprint. Sections \ref{SpecClass} and \ref{results} describe our spectral classification method and results, respectively. Finally, we summarize our results and conclusions in Section \ref{discussion}.

Throughout this work we use h = 0.70, $\Omega_m$ = 0.30, and $\Omega_{\Lambda}$ = 0.70, where the Hubble constant is $H_0 = 100$ h km s$^{-1}$ Mpc$^{-1}$ \citep{tegmark2004cosmological}. Spectral indices are defined by flux density $S_{\nu} \propto \nu^{-\alpha}$, indicating a flat spectrum when $\alpha < 0.5$. The WISE magnitudes are in the Vega system and are not corrected for the Galactic extinction, since, as shown in \cite{dAbrusco2014wibrals}, such correction only affects the magnitude at 3.4 $\mu$m for sources lying close to the Galactic plane and it ranges between 2\% and 5\% of the magnitude, thus not affecting significantly the results. WISE bands are indicated as W1, W1, W3 and W4, and correspond respectively to the nominal wavelengths at 3.4, 4.6, 12, and 22 $\mu$m.

\begin{figure}
    \centering
    \includegraphics[scale=0.5]{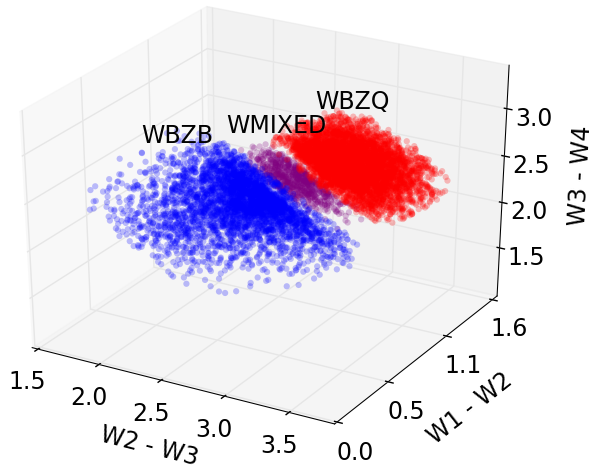}
    \caption{The whole WIBRaLS2 sample divided into three categories accordingly to the sources' IR colors: BZB candidates (WBZBs, in blue), BZQ candidates (WBZQs, in red) and blazars with intermediate colors (WMIXEDs, in purple).}
    \label{fig:classes}
\end{figure}


\section{Sample selection}
\label{SampleSelection}

For this analysis we used observations available in the AllWISE catalog \citep{cutri2013explanatory,cutri2014vizier}, which contains astrometry and photometry in the IR for $\sim $3$ \times $10$^7$ sources in W1, W2, W3 ad W4 bands, it includes data from the cryogenic and post-cryogenic survey phases \citep{mainzer2011neowise}, increasing to $\sim 10^8$ sources detected only on the first three bands W1, W2 and W3.

\subsection{WIBRaLS2}
\label{Sample:WIBRaLS}

\begin{figure}
    \centering
    \includegraphics[scale=0.6]{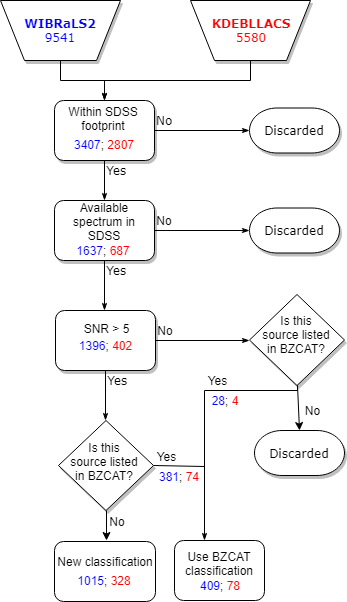}
    \caption{Flow chart with all steps adopted in the sample selection and spectral classification. The numbers in blue refer to sources in WIBRaLS2 while numbers in red refer to sources in KDEBLLACS after each step.}
    \label{fig:flowchart}
\end{figure}

\begin{figure*}
    \centering
    \includegraphics[scale=0.4]{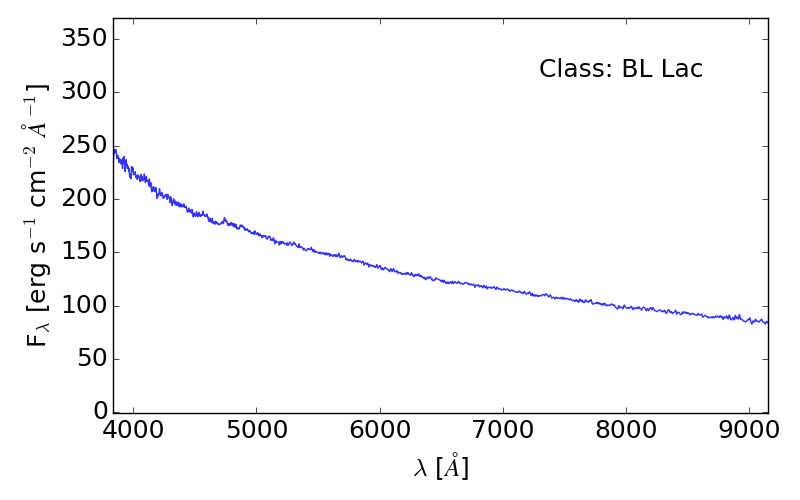}
    \includegraphics[scale=0.4]{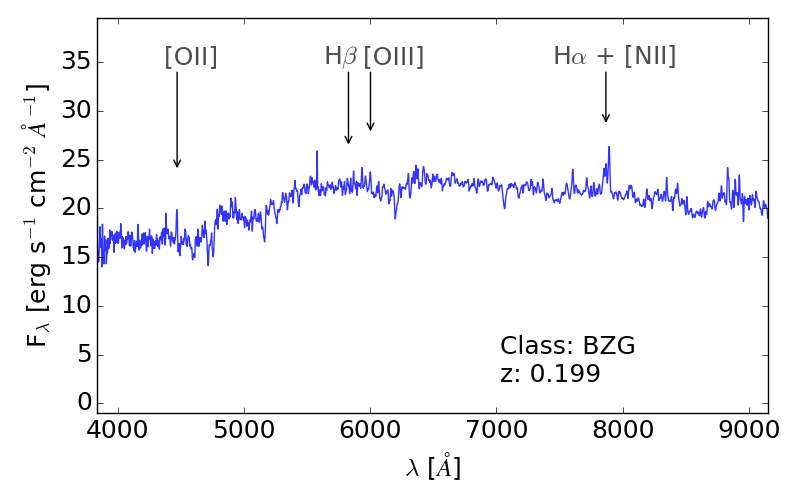}
    \includegraphics[scale=0.4]{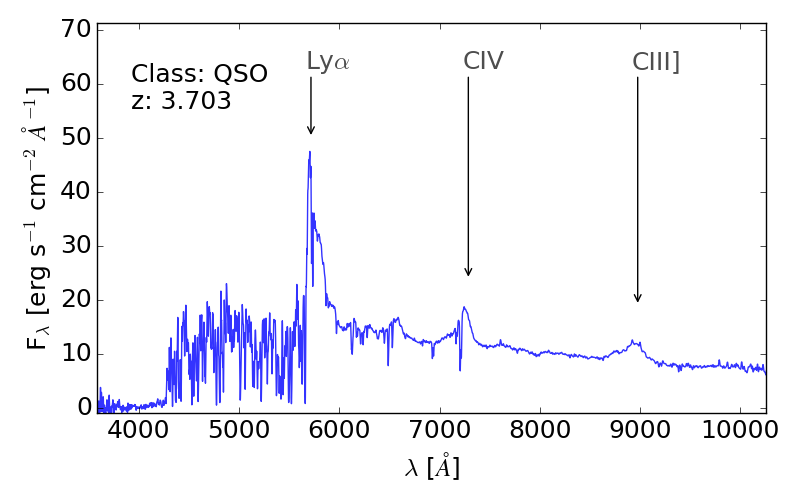}
    \includegraphics[scale=0.4]{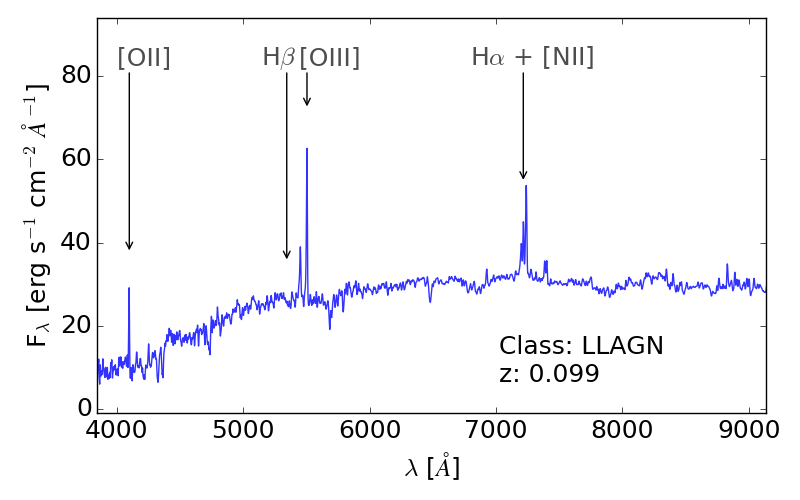}
    \includegraphics[scale=0.4]{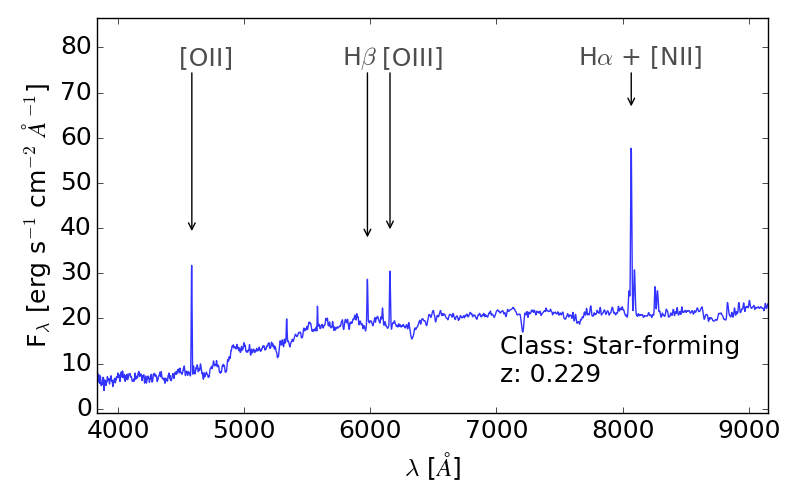}
    \includegraphics[scale=0.4]{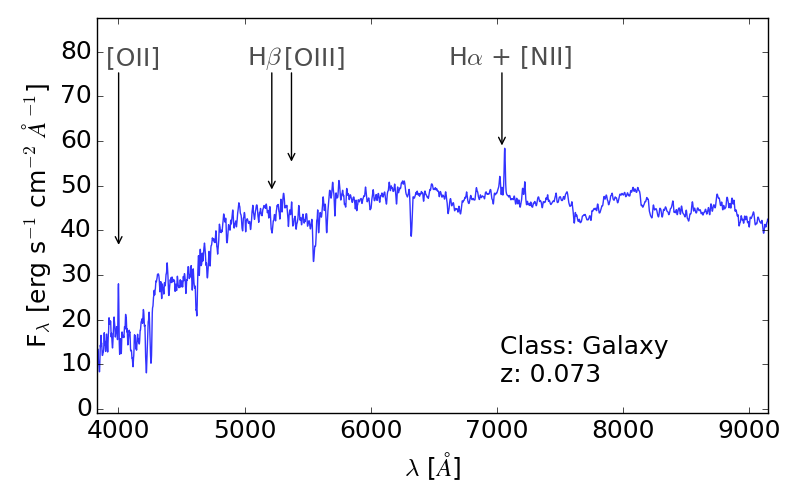}
    \caption{Typical smoothed spectra in observed wavelengths for each one of the spectral categories found in WIBRaLS2 and KDEBLLACS.}
    \label{fig:spectra}
\end{figure*}

The WIBRaLS2 catalog was conceived to provide a good sample of potential $\gamma$-ray sources based on IR and radio data. It includes mid-IR sources detected in the four WISE filters, with colors similar to those of blazars listed in the \textit{Fermi}-LAT 4-year Point Source Catalog \citep[3FGL;][]{acero2015_3FGL}, spatially cross-matched with a radio counterpart found in one of the three major radio surveys: the National Radio Astronomy Observatories Very Large Array (VLA) Sky Survey \citep[NVSS;][]{condon1998nrao}, the VLA Faint Images of the Radio Sky at Twenty-cm Survey \citep[FIRST;][]{white1997catalog,helfand2015last} and the Sydney University Molonglo Sky Survey Source Catalog \citep[SUMSS;][]{mauch2003sumss}; and selected to be radio-loud based on their $q_{22}$ spectral parameter, defined as $q_{22} = \log(S_{22\mu m}/S_{radio})$, with $S_{22\mu m}$ and $S_{radio}$ being the flux densities at $22\mu m$ and in radio respectively. The total number of blazar candidates in WIBRaLS2 is 9541 and they are classified (see Figure \ref{fig:classes}) as candidate BZBs, candidate BZQs or MIXEDs (hereinafter WBZBs, WBZQs and WMIXEDs, respectively), the latter defined as those with mid-IR colors which are intermediate between WBZBs and WBZQs \citep[for more details, see][]{dAbrusco2019wibrals2_KDEBLLACS}.

In this paper, we have cross-matched the WIBRaLS2 catalog with SDSS DR15 associating blazar candidates listed therein with their optical counterparts within an angular separation of $2''$ \citep{massaro2014optical}. We found a total number of 3407 unique associations and then used the SDSS Science Archive Server\footnote{\url{https://dr15.sdss.org/optical/spectrum/search}} to select only those sources with an available optical spectrum. We then selected only those sources with a spectral signal to noise ratio (SNR) above 5 and sources with SNR < 5 but listed in Roma-BZCAT. This led us to a final sample of 1424 WIBRaLS2 sources, which corresponds to $\sim 15\%$ of the whole WIBRaLS2. A flowchart describing the effects of our selection of the number of candidates in the WIBRaLS2 catalog is displayed in Figure~\ref{fig:flowchart}.

\subsection{KDEBLLACS}
\label{Sample:KDE}

The KDEBLLACS catalog was built by first applying a kernel density estimation (KDE) to a two-dimensional distribution of training set sources of BZB type in the WISE W1-W2 $\times$ W2-W3 color diagram to determine its probability distribution function (PDF). Then sources were selected when located within the 90\% isodensity contour of the training set \citep{dAbrusco2019wibrals2_KDEBLLACS} built with the KDE. The color uncertainty ellipses of each source in KDEBLLACS must be fully contained in this 90\% contour. Sources in KDEBLLACS are  radio-loud according to the  $q_{12}$ parameter, defined as $q_{12} = \log(S_{12\mu m}/S_{radio})$, with $S_{12\mu m}$ and $S_{radio}$ being the flux densities at $12\mu m$ and in radio respectively \citep{dAbrusco2019wibrals2_KDEBLLACS}.

KDEBLLACS lists 5580 candidate BZBs (hereinafter labelled as KBZBs). This number decreases to 2807 sources lying in the SDSS  footprint, however, only 402 of them have good (i.e., SNR > 5) optical spectra. Our final sample consists of these 402 sources together with 4 sources listed in Roma-BZCAT which have low significance (SNR < 5) SDSS spectra. This sample corresponds to 7\% of the whole KDEBLLACS (see Figure~\ref{fig:flowchart} for details).

\section{Spectral analysis and classification}
\label{SpecClass}

To classify sources in WIBRaLS2 and KDEBLLACS according to their optical spectra, we started by cross-matching both samples, previously defined, with Roma-BZCAT \citep{massaro2015romabzcat5th}, finding a total of 409 counterparts (out of 1424) for WIBRaLS2 and 78 (out of 404) for KDEBLLACS, respectively. These sources were already classified as blazars and we assumed their classification without further check. 

We then adopted the following classification scheme, divided into sources listed in Roma-BZCAT and remaining sources:

\begin{itemize}
    \item Roma-BZCAT
          \begin{itemize}
              \item BZB
              \item BL Lac - galaxy-dominated (BZG)
              \item Blazar uncertain type (BZU)
              \item BZQ
          \end{itemize}
    
    \item Remaining sources 
          \begin{itemize}
              \item BZB
              \item Quasi-stellar object (QSO), or BZQ in some cases
              \item Low-luminosity active galactic nucleus (LLAGN)
              \item Star-forming galaxy
              \item BZG
              \item Galaxy
              \item Star
         \end{itemize}

\end{itemize}
Figure \ref{fig:spectra} shows the typical smoothed spectra for all classes discussed above. The criteria adopted in the classification of the new analyzed sources is explained below.

Particularly for blazars, we classified sources as BZBs when they showed a featureless optical spectra with a dominant blue continuum, and as QSOs when presenting redshifted broad emission lines again above a dominant blue continuum coupled with a radio luminosity $L_r$ greater than 10$^{38}$ erg s$^{-1}$.

\begin{figure*}
    \centering
    \includegraphics[width=\linewidth]{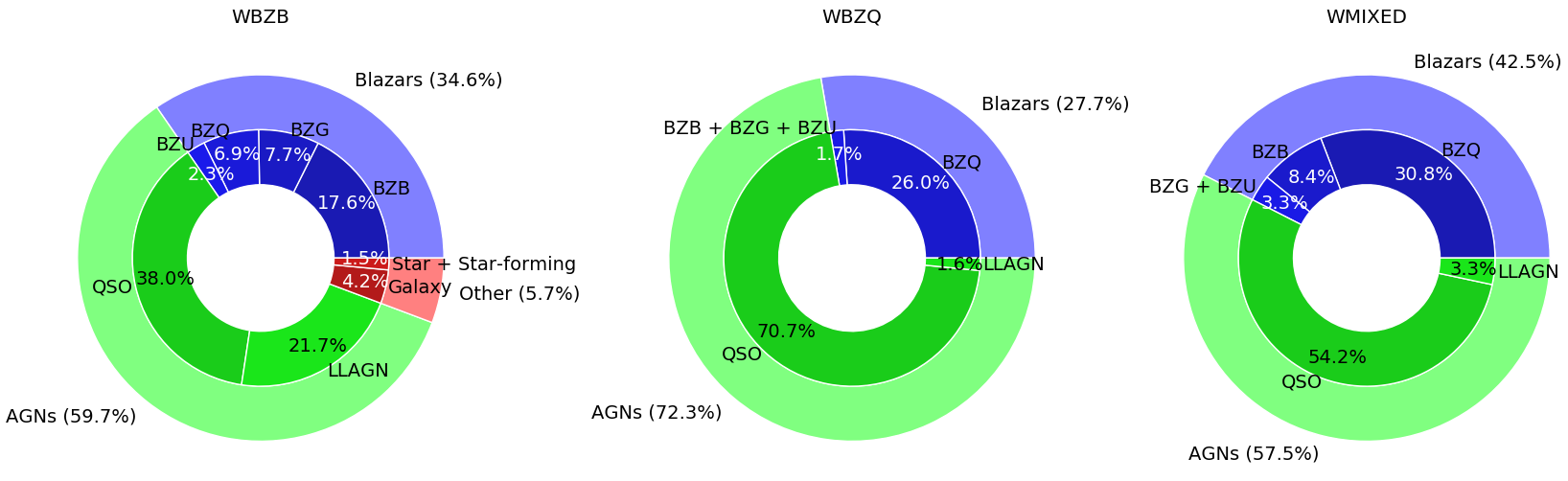}
    \caption{Nested pie charts for the 3 classes of sources described in WIBRaLS2 (WBZB, WBZQ and WMIXED). The major contaminant in all categories are AGNs. A significant fraction of these AGNs, however, may be blazars (BZQ). The subclass LLAGN is represented by Seyferts and LINERs.}
    \label{fig:pieWIBRALS}
\end{figure*}

\begin{figure*}
    \centering
    \includegraphics[scale=0.3]{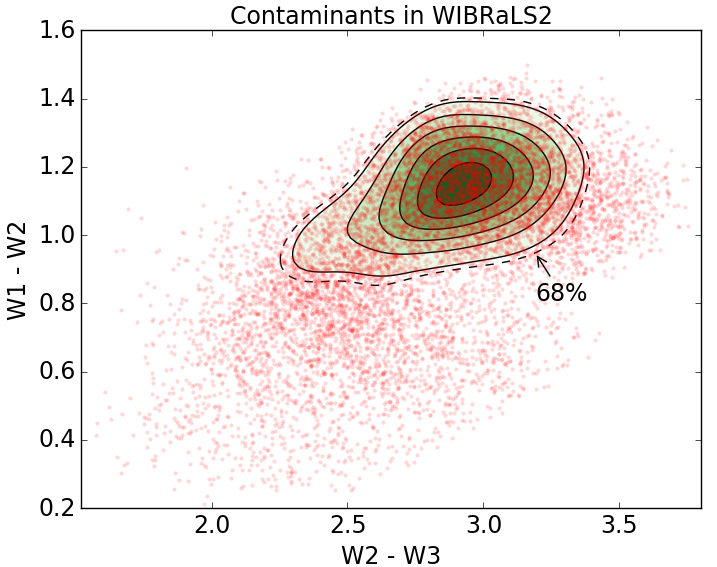}
    \includegraphics[scale=0.3]{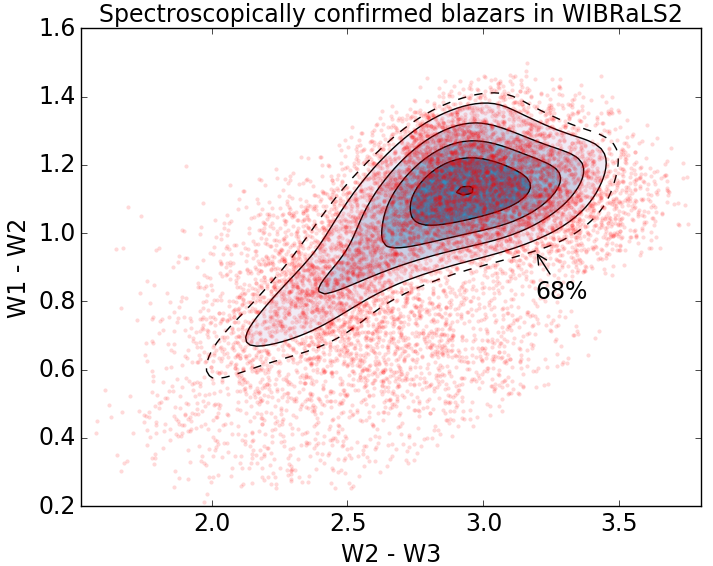}
    \includegraphics[scale=0.3]{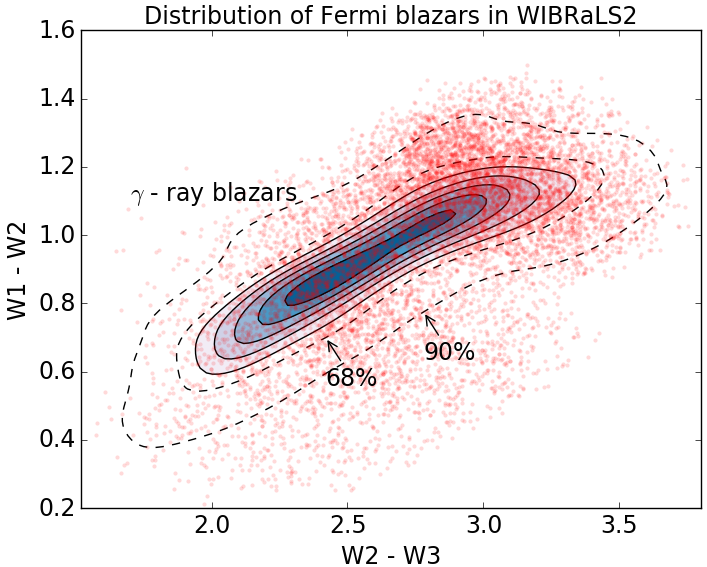}
    \caption{Contaminants in WIBRaLS2. Left: Contamination of WIBRaLS2 obtained based on a KDE and using the SDSS DR15 spectra of 953 sources (833 QSOs, 100 Seyferts and 20 LINERs) as training sample. The contaminants are mainly QSOs concentrated in the WBZQ region of the color-color diagram (as pointed in Figure \ref{fig:classes}). Middle: PDF of spectroscopically confirmed blazars in WIBRaLS2 (KDE training set composed of 281 BZQs, 95 BZBs, 36 BZGs and 24 BZUs). Most of them lie in the most contaminated region of WIBRaLS2. Right: PDF of $\gamma$-ray blazars obtained using 1320 cross-matches between 4FGL blazars and WIBRaLS2 sources as a training sample for a KDE. The $\gamma$-ray blazars are mainly located in a narrow strip peaking in the WBZB region. The solid lines represent the isodensity contours obtained with the KDE, while the dashed lines indicate the 68\% and (for the last panel) 90\%.}
    \label{fig:Contamination_WIBRaLS2}
\end{figure*}

On the other hand, LLAGNs and star-forming galaxies were distinguished based on their line ratios [OIII]/H$_{\beta}$ and [NII]/H$_{\alpha}$ according to the BPT diagram \citep{baldwin1981_BPT,kewley2006host}. A total of 4 star-forming galaxies were identified in WIBRaLS2 and only 2 in KDEBLLACS.

To distinguish between normal galaxies and BZGs, we measured their relative flux depression bluewards the Ca II break. This parameter was defined as $CB = (F_+ - F_-)/F_+$, with $F_+$ and $F_-$ meaning the flux densities measured in ranges of 200 \AA $ $ at wavelengths just above and below that of the Ca II H\&K break \citep{dressler1987systematics,massaro2012colours}. As in \cite{stocke1991einstein}, we adopted a threshold value of 0.25 for CB to ensure the presence of a substantial non-thermal continuum (i.e., CB$<$0.25) leading to a BZG classification; otherwise, it was classified as a normal galaxy. 

Finally, a total of 3 and 2 stars were found contaminating WIBRaLS2 and KDEBLLACS respectively; four of them typical cold M- or G-type, and one white dwarf.


\section{Results}
\label{results}

The optical classification previously described enabled us to characterize WIBRaLS2 and KDEBLLACS subsamples. The results are presented here splitted in the following subsections.

\subsection{WIBRALS2}
\label{WIBRaLS_Results}

We analyzed a total of 1424 spectra (1396 with SNR > 5 and 28 with SNR < 5 but with a counterpart in Roma-BZCAT) among the 9541 sources available in WIBRaLS2 -- but only 3407 in the SDSS footprint. According to their WIBRaLS2 classification, these sources were divided into 471 WBZBs, 833 WBZQs and 120 WMIXEDs. As can be seen in Figure \ref{fig:pieWIBRALS}, their main contaminant class are QSOs. A significant fraction of these QSOs, however, may have a radio flat spectrum, indicating that they could be indeed blazars of BZQ type. Then, we found that 17.6\% of WBZBs show a featureless optical spectrum; 26.1\% of WBZQs are confirmed BZQs (i.e., radio flat spectrum) and 30.8\% of WMIXEDs sources are also BZQs.

As can be seen in the left and middle panels of Figure \ref{fig:Contamination_WIBRaLS2}, the majority of contaminants in WIBRaLS2 are concentrated in the WBZQ and WMIXED regions of the mid-IR color-color diagram. On the other hand, the WBZB sample is cleaner, and its purity can reach $\sim 50\%$ in the area bounded by the 68\% 
isodensity contours of 4FGL $\gamma$-ray blazars (Figure \ref{fig:Contamination_WIBRaLS2}, right panel).

\subsection{KDEBLLACS}
\label{KDE_Results}

\begin{figure}
    \centering
    \includegraphics[width=\linewidth]{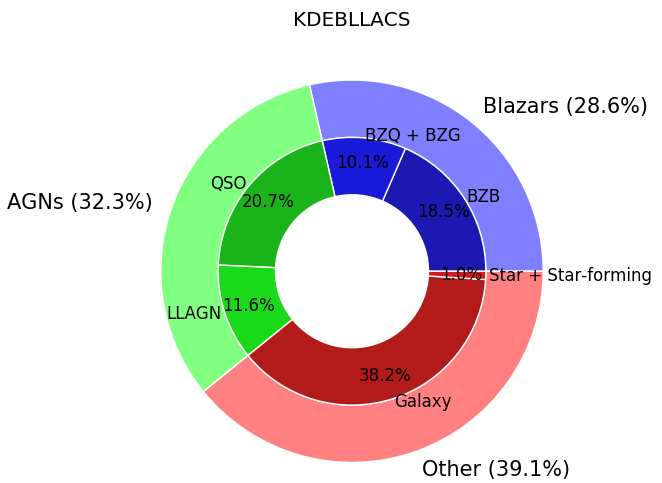}
    \caption{Contaminants for KDEBLLACS divided by classes. The major contaminant for this catalog are normal galaxies. Some fraction of the QSOs can have a flat radio spectrum and be BZQs.}
    \label{fig:pieKDE}
\end{figure}

With a total of 406 spectra analyzed, KBZBs in KDEBLLACS were classified as follows: 75 BZBs (60 of them confirmed in Roma-BZCAT), 39 BZGs (17 confirmed in Roma-BZCAT), 155 normal galaxies, 47 LLAGNs (20 LINERs and 27 Seyferts), 84 QSOs, 2 confirmed BZQs (1 of them listed in Roma-BZCAT), 2 stars and 2 star-forming galaxies, all according to the criteria previously described. Figure \ref{fig:pieKDE} summarizes these results in a nested pie chart.

\begin{figure*}
    \centering
    \includegraphics[scale=0.45]{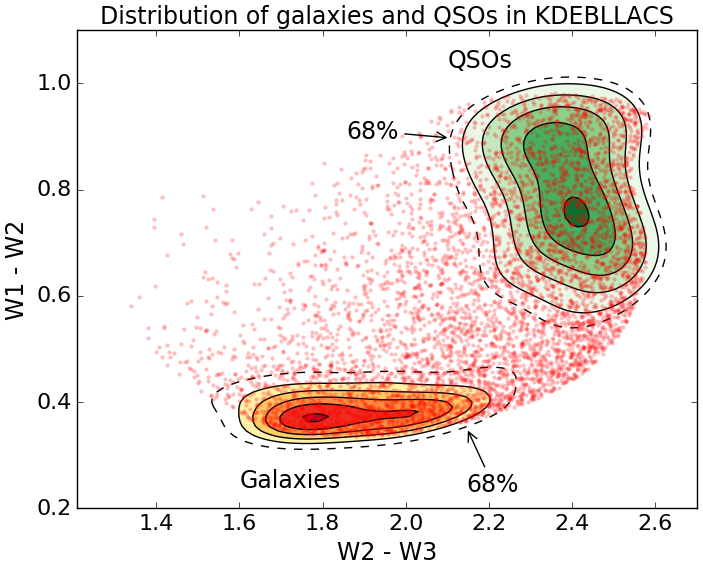}
    \includegraphics[scale=0.45]{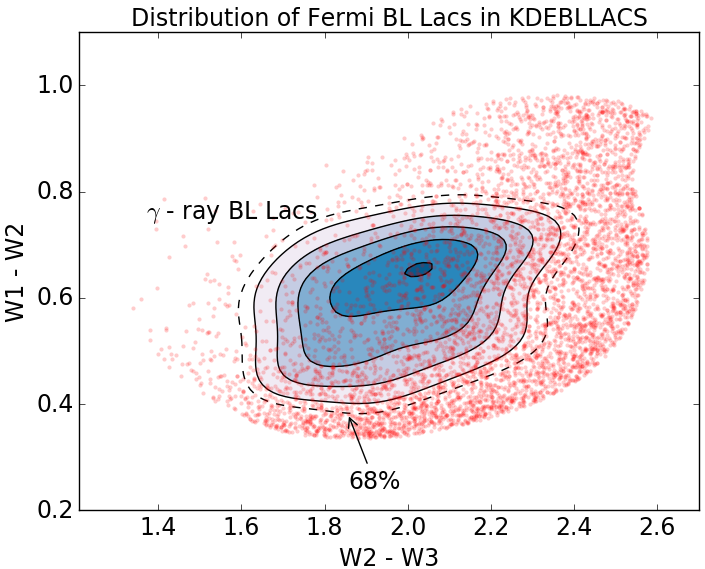}
    \caption{Contamination of KDEBLLACS. Left: The main contaminants of KDEBLLACS are normal galaxies and QSOs, both located in the edges of the sample. Right: PDF of the $\gamma$-ray BZBs available in 4FGL obtained with a KDE. We can see that the majority of BZBs are indeed located outside of the two contaminated zones. The solid lines represent the KDE isodensity contours, while the dashed lines represents the 68\% isodensity contour.}
    \label{fig:Contamination_KDE}
\end{figure*}

It is worth noting that normal galaxies and QSOs are the major contaminant classes of KDEBLLACS selection criteria. After applying two KDEs to this sample, one using spectroscopically confirmed QSOs and the other using spectroscopically confirmed galaxies as training sets, we observe (as expected) that these contaminants are concentrated towards the edges of the mid-IR color-color diagram (Figure \ref{fig:Contamination_KDE}, left panel), mainly in the bottom and upper-right corners. Indeed, most of the BZBs classified in 4FGL lie outside of these regions (Figure \ref{fig:Contamination_KDE}, right panel).

\subsection{Selection effects}
\label{subSec:selection_effects}

\begin{figure}
    \centering
    \includegraphics[width=\linewidth]{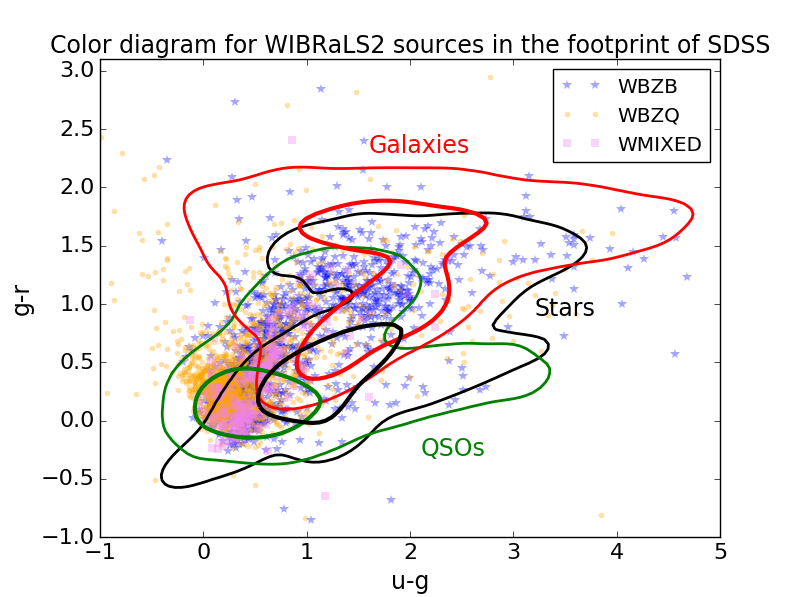}
    \includegraphics[width=\linewidth]{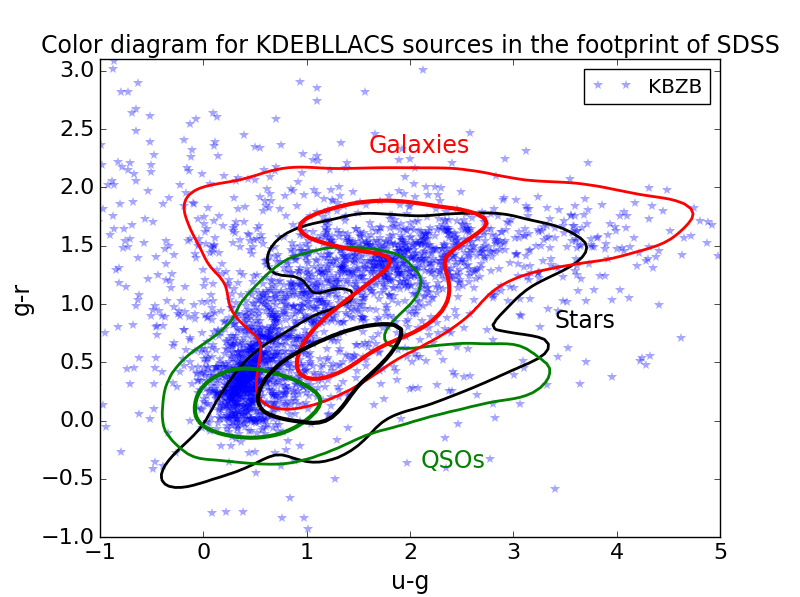}
    \caption{Optical color-color diagrams for WIBRaLS2 (upper panel) and KDEBLLACS (bottom panel) sources in the footprint of SDSS. The total number of sources is 3407 for WIBRaLS2 and 2807 for KDEBLLACS, and are divided in WBZBs/KBZBs (blue stars), WBZQs (orange circles) and WMIXEDs (violet squares), accordingly to their classification in WIBRaLS2 and KDEBLLACS. The red, green and black thin and thick lines represent, respectively, the 90\% and 50\% isodensity contours based on the distribution of 10\,000 Galaxies, 10\,000 QSOs and 10\,000 Stars with available spectra in SDSS DR15.}
    \label{fig:Color-color_SDSS}
\end{figure}

Since $\sim\!52\%$ of WIBRaLS2 and $\sim\!75\%$ of KDEBLLACS sources in the footprint of SDSS have no available spectrum, we investigated possible 
selection effects, starting by comparing the typical optical colors of sources in WIBRaLS2 and KDEBLLACS with the colors of sources spectroscopically 
observed by SDSS, split by their SDSS spectroscopic classification. The lines in Figure \ref{fig:Color-color_SDSS} show the 90\% and 50\% isodensity contours 
for 10\,000 spectroscopically classified Galaxies (red), QSOs (green) and Stars (black), randomly selected from SDSS DR15. Overall, the candidates from WIBRaLS2 and KDEBLLACS catalog occupy very similar region of the $u$-$g$ vs $g$-$r$ optical diagram (Figure~\ref{fig:Color-color_SDSS}). Most of the WIBRaLS2 and KDEBLLACS candidates are located within the 90\% isodensity contours defined by spectroscopic SDSS QSOs and Galaxies, which significantly overlap with the 90\% contour for stars. In particular, in the case of WIBRaLS2 candidates (top 
panel), a large fraction (48\%) of candidates lie within the 50\% contour of sources classified as QSOs, although other sources are scattered in the Galaxies and Stars-dominated areas. WIBRaLS2 candidates classified as WBZQ unsurpisingly peak in the area occupied by spectroscopically confirmed SDSS QSOs, while intermediate (WMIXED) and WBZB sources spread a much larger area consistent with both SDSS spectroscopic Stars and Galaxies. KDEBLLACS candidates are more evenly scattered through the isodensity contours for all three spectroscopic classes, indicating a likely larger contamination from the host galaxies. Similar behaviors are visible in the other SDSS color-color diagrams, not reported in Figure~\ref{fig:Color-color_SDSS}. In general, we can rule out the existence of significant selection effects due to the colors of the optical counterparts of our candidates.

\begin{figure}
    \centering
    \includegraphics[width=\linewidth]{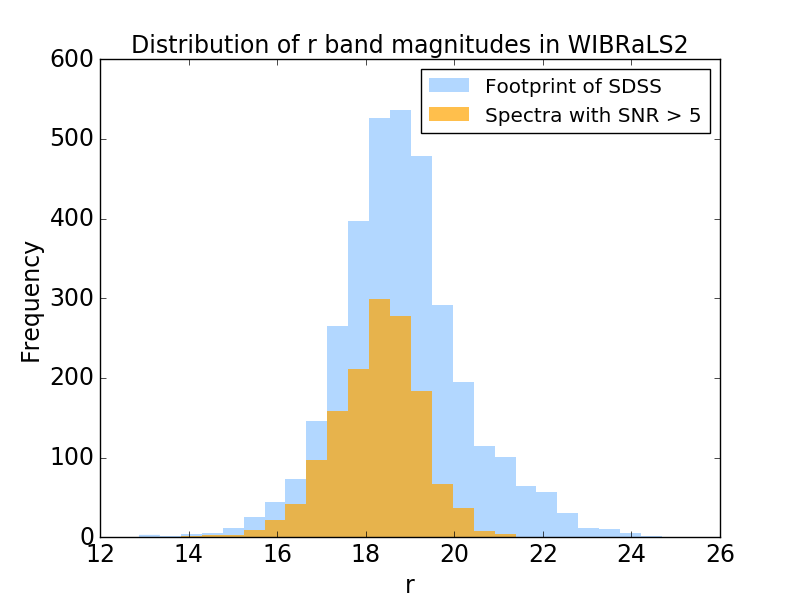}
    \includegraphics[width=\linewidth]{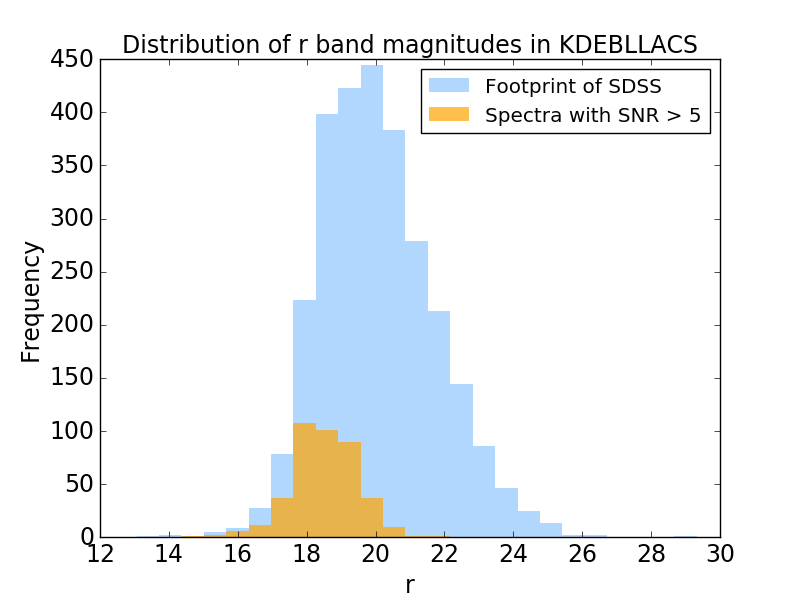}
    \caption{Distribution of r band magnitudes for WIBRaLS2 (upper panel) and KDEBLLACS (bottom panel). The blue histograms stand for all WIBRaLS2 and KDEBLLACS sources in the footprint of SDSS (3407 and 2807, respectively), while the orange histograms are only for sources with high quality (SNR > 5) spectra in SDSS DR15.}
    \label{fig:r_distribution_SDSS}
\end{figure}

Selection effects caused by the magnitude distribution of our candidates, however, are of major importance. As shown in 
Figure~\ref{fig:r_distribution_SDSS}, a large majority of the candidates missing optical spectra, especially for KDEBLLACS catalog, are fainter than the candidates that have been followed-up spectroscopically. In particular, the reason why KDEBLLACS is spectroscopically less complete than WIBRaLS2 is because this catalog has in general fainter sources. This effect is expected as KDEBLLACS candidates appear to be 
intrinsically fainter than WIBRaLS2 sources in all three WISE bands (W1, W2 and W3) where they are detected, as shown in Figure 7 from~\citet{dAbrusco2019wibrals2_KDEBLLACS}. The magnitude distributions of WIBRaLS2 candidates with and without SDSS spectroscopic counterparts (upper panel in Figure~\ref{fig:r_distribution_SDSS}), are very similar. This suggests that other selection effects, likely based on the colors, extension and morphology of the sources, increase the number of WIBRaLS2 sources lacking  SDSS spectroscopic counterparts\footnote{For a general picture of the SDSS selection criteria for spectroscopic observations, see \url{https://www.sdss.org/dr15/algorithms/legacy_target_selection/}}.

\subsection{Blazar candidates of uncertain type}

Some sources in the \textit{Fermi}-LAT catalogs are considered blazar candidates of uncertain type (BCUs) because the adopted association methods select a counterpart that satisfies at least one of the following conditions \citep{ackermann2015_3LAC,lat20194fgl}: i) An object classified as blazar of uncertain or transitional type in Roma-BZCAT. ii) A source with multiwavelength data indicating a typical two-humped blazar-like spectral energy distribution (SED) and/or a flat radio spectrum. BCUs are divided into three sub-types \citep{ackermann2015_3LAC}:

\begin{itemize}
    \item BCU I: the counterpart has a published optical spectrum which is not sensitive enough for classifying it as BZQ or BZB.
    \item BCU II: there is no available optical spectrum but an evaluation of the SED synchrotron peak position is possible.
    \item BCU III: the counterpart shows typical blazar broadband emission and a flat radio spectrum, but lacks a optical spectrum and reliable measurement of the synchrotron peak position. 
\end{itemize}

In 4FGL, 1155 sources are considered as BCUs. Our analysis based on the optical spectra available in SDSS DR15 (Section \ref{SpecClass}) allowed us to give a conclusive classification for 11 of them, as shown in Table \ref{tab:BCUs_WIBRaLS_and_KDE}.

The 4FGL catalog associated the source 4FGL J0038.7-0204 with the broad line radio galaxy (RDG) 3C\,17 at redshift $\sim$0.22 as previously determined by \cite{schmidt1965optical} in agreement with the SDSS spectrum we analyzed. The kiloparsec-scale radio morphology of 3C\,17 is dominated by a single-sided, curved jet \citep{morganti1993radio} also shining in the X-rays \citep{massaro2009jet,massaro2010chandra}. Recent optical spectroscopic observations also revealed that 3C\,17 is the brightest cluster galaxy in its large scale environment \citep{madrid20183c}.

\begin{table}
    \centering
    \begin{tabular}{c|c|c|c}
        WISE name & 4FGL name & Class & z \\
        \hline
        J003719.15+261312.6	& J0037.9+2612 & BZG & 0.1477\\
        J003820.53-020740.5	& J0038.7-0204 & RDG & 0.2204\\
        J013859.14+260015.7	& J0139.0+2601 & BZB & --\\
        J020239.94-030207.9	& J0202.6-0258 & BZQ & 1.3444\\
        J084734.29+460928.0	& J0846.9+4608 & BZQ & 1.2165\\
        J094452.09+520233.4	& J0945.2+5200 & BZQ & 0.5630\\
        J095608.57+393515.8	& J0956.0+3936 & BZQ & 1.1730\\
        J130407.31+370908.1	& J1304.0+3704 & BZB & --\\
        J134243.61+050432.1	& J1342.7+0505 & BZG & 0.1365\\
        \hline
        J123124.08+371102.2	& J1230.9+3711 & BZG & 0.2180\\
        J154150.09+141437.6	& J1541.7+1413 & BZG & 0.2230
    \end{tabular}
    \caption{Spectral classification for 11 sources classified as BCUs in 4FGL. This table is segmented in WIBRaLS2 (upper part) and KDEBLLACS (bottom). The columns give the name of the sources as in WISE and 4FGL, our optical classification and the redshifts as in SDSS DR15.}
    \label{tab:BCUs_WIBRaLS_and_KDE}
\end{table}


\subsection{New BZBs}

During the spectroscopic classification (Section \ref{SpecClass}), we found a total of 25 (10 in WIBRaLS2 and 15 in KDEBLLACS) sources with featureless optical spectra and a relatively strong blue continuum -- typical characteristics of BZB non-thermal emission -- which are not available in Roma-BZCAT. Their smoothed spectra, names/positions and r band magnitudes can be found in the Appendix \ref{sec.foo}.

To check these new BZBs, we considered the $u-r$ color index of the SDSS photometric system (AB magnitudes), which is an efficient discriminator between BZBs and nuclei with weak or even absent activity \citep{massaro2012colours}. When computing these indexes, we considered the extinction-corrected SDSS model magnitudes accordingly to the following formula:
\begin{equation}
    (u - r) = (u - r)_{obs} - 0.81 A_r
\end{equation}
where $A_r$ is the extinction in the r band, given in the SDSS database. Typically, BZBs present a color index $u - r \leq 1.4$ \citep{massaro2012colours}, however, this criterium is not suited for sources at high redshifts (z > 0.5). Table \ref{tab:new_Bllacs} summarizes our findings.

\begin{table}
    \centering
    \begin{tabular}{c|c|c}
        WISE name & SDSS name & $u - r$ \\
        \hline
        J013859.14+260015.7 & J013859.14+260015.7 & 1.09\\
        J085446.24-003348.1 & J085446.22-003349.5 & 0.82\\
        J093522.08+502932.2 & J093522.08+502932.2 & 0.88\\
        J115406.13+185723.6 & J115406.11+185723.6 & 1.16\\
        J130407.31+370908.1 & J130407.32+370908.1 & 1.11\\
        J144906.05+071701.2 & J144906.04+071701.3 & 0.86\\
        J165558.59+391218.1 & J165558.61+391218.2 & 0.91\\
        J215601.64+181837.1 & J215601.64+181837.0 & 0.45\\
        J220812.70+035304.5 & J220812.70+035304.6 & 0.85\\
        J235915.62+221450.0 & J235915.62+221450.1 & 1.38\\
        \hline
        J000710.65-032029.4 & J000710.65-032029.4 & 0.74\\
        J020303.61-024547.3 & J020303.62-024546.9 & 0.82\\
        J024024.36-025334.3 & J024024.37-025334.3 & 0.70\\
        J083706.00+583152.9 & J083706.01+583152.9 & 0.60\\
        J083955.10+121702.9 & J083955.09+121703.0 & 0.71\\
        J091804.13+071653.6 & J091804.13+071653.6 & 1.22\\
        J101147.54+360018.7 & J101147.50+360018.8 & 0.68\\
        J103852.20+325651.7 & J103852.17+325651.6 & 1.22\\
        J111356.24+552255.3 & J111356.30+552255.6 & 0.60\\
        J114221.77+334201.8 & J114221.76+334201.8 & 0.85\\
        J114352.67+155821.9 & J114352.66+155822.0 & 0.51\\
        J135154.45+285008.0 & J135154.45+285007.9 & 1.26\\
        J154714.91+265442.2 & J154714.91+265442.4 & 0.77\\
        J171747.84+392607.3 & J171747.85+392607.3 & 0.99\\
        J171841.42+360522.6 & J171841.44+360522.2 & 1.32
    \end{tabular}
    \caption{New BZBs found during the spectroscopic classification in Section \ref{SpecClass}. The first and second columns present the name of the objects in WISE and SDSS. The last column shows the color index $u-r$ based on SDSS AB magnitudes. The upper section of the table shows the BZBs found in WIBRaLS2 while the bottom section shows the ones found in KDEBLLACS.}
    \label{tab:new_Bllacs}
\end{table}


\section{Summary and conclusions}
\label{discussion}

In this work we characterized the two newly released catalogs of blazar candidates WIBRaLS2 and KDEBLLACS based on 1830 optical spectra -- 1798 spectra with SNR > 5 and 32 with SNR < 5 but with a counterpart in Roma-BZCAT -- available in SDSS DR15 data. Both catalogs indeed presented a high number of spectroscopic confirmed blazars, although the contamination level of WIBRaLS2 was $\sim 60\%$, mainly due to QSOs, and the contamination of KDEBLLACS was $\sim 70\%$, mainly due to QSOs and normal galaxies. We stress that these estimates of the contamination represent upper limits, 
due to the selection effects affecting the SDSS spectroscopic sample used in this paper, as described in Section~\ref{subSec:selection_effects}. 
Main results can be summarized as follows:

\begin{itemize}
    \item Including the objects in Roma-BZCAT, 34.6\%, 27.7\% and 42.5\% of the sources respectively in WBZB, WBZQ and WMIXED are blazars, which gives an overall weighted lower limit to the efficiency of WIBRaLS2 blazar selection of $\approx 31\%$, with the major contaminants being, as expected, QSOs. 
    \item The lower limits to the efficiency of KDEBLLACS in identifying blazars is $\sim 30\%$, and its contaminants -- mainly galaxies and QSOs -- are concentrated on the edges of the mid-IR color-color diagram (Figure \ref{fig:Contamination_KDE}).
    \item The spectral analysis carried on in Section \ref{SpecClass} led us to the discovery of 25 new BZBs not available anywhere else in literature and to the classification of 11 BCUs listed in 4FGL.
\end{itemize}

This work contributes to a better understanding of the $\gamma$-ray sky in the \textit{Fermi}-LAT era. In particular, the community will benefit from the characterization of WIBRaLS2 and KDEBLLACS in population studies of blazars and in subsequent programs of spectroscopic follow-up needed to confirm the nature of the UGSs.

\begin{acknowledgements}
      We thank the anonymous referee for constructive comments which helped to improve the manuscript. The accomplishment of this project was only possible due to the ongoing support from the S\~ao Paulo Research Foundation (FAPESP), grants 2016/25484-9, 2018/24801-6 and 2017/01461-2. PH acknowledges support from the CONACyT program for Ph.D. studies. FR acknowledges support from FONDECYT Postdoctorado 3180506 and CONICYT Chile grant Basal-CATA PFB-06/2007. R.D'A. is supported by NASA contract NAS8-03060 (Chandra X-ray Center).

This work is supported by the ``Departments of Excellence 2018 - 2022'' Grant awarded by the Italian Ministry of Education, University and Research (MIUR) (L. 232/2016). This research has made use of resources provided by the Compagnia di San Paolo for the grant awarded on the BLENV project (S1618\_L1\_MASF\_01) and by the Ministry of Education, Universities and Research for the grant MASF\_FFABR\_17\_01. F.M. acknowledges financial contribution from the agreement ASI-INAF n.2017-14-H.0. A.P. acknowledges financial support from the Consorzio Interuniversitario per la fisica Spaziale (CIFS) under the agreement related to the grant MASF\_CONTR\_FIN\_18\_02.

TOPCAT8 \citep{taylor2005topcat} was extensively used in this work for the preparation and manipulation of the tabular data. This publication makes use of data products from the \textit{Wide-field Infrared Survey Explorer}, which is a joint project of the University of California, Los Angeles, and the Jet Propulsion Laboratory/California Institute of Technology, funded by the National Aeronautics and Space Administration.

Funding for the Sloan Digital Sky Survey IV has been provided by the Alfred P. Sloan Foundation, the U.S. Department of Energy Office of Science, and the Participating Institutions. SDSS-IV acknowledges
support and resources from the Center for High-Performance Computing at
the University of Utah. The SDSS web site is \url{www.sdss.org}.

SDSS-IV is managed by the Astrophysical Research Consortium for the 
Participating Institutions of the SDSS Collaboration including the 
Brazilian Participation Group, the Carnegie Institution for Science, 
Carnegie Mellon University, the Chilean Participation Group, the French Participation Group, Harvard-Smithsonian Center for Astrophysics, 
Instituto de Astrof\'isica de Canarias, The Johns Hopkins University, Kavli Institute for the Physics and Mathematics of the Universe (IPMU) / 
University of Tokyo, the Korean Participation Group, Lawrence Berkeley National Laboratory, 
Leibniz Institut f\"ur Astrophysik Potsdam (AIP),  
Max-Planck-Institut f\"ur Astronomie (MPIA Heidelberg), 
Max-Planck-Institut f\"ur Astrophysik (MPA Garching), 
Max-Planck-Institut f\"ur Extraterrestrische Physik (MPE), 
National Astronomical Observatories of China, New Mexico State University, 
New York University, University of Notre Dame, 
Observat\'ario Nacional / MCTI, The Ohio State University, 
Pennsylvania State University, Shanghai Astronomical Observatory, 
United Kingdom Participation Group,
Universidad Nacional Aut\'onoma de M\'exico, University of Arizona, 
University of Colorado Boulder, University of Oxford, University of Portsmouth, 
University of Utah, University of Virginia, University of Washington, University of Wisconsin, 
Vanderbilt University, and Yale University.

The \textit{Fermi} LAT Collaboration acknowledges generous ongoing support
from a number of agencies and institutes that have supported both the
development and the operation of the LAT as well as scientific data analysis.
These include the National Aeronautics and Space Administration and the
Department of Energy in the United States, the Commissariat \`a l'Energie Atomique
and the Centre National de la Recherche Scientifique / Institut National de Physique
Nucl\'eaire et de Physique des Particules in France, the Agenzia Spaziale Italiana
and the Istituto Nazionale di Fisica Nucleare in Italy, the Ministry of Education,
Culture, Sports, Science and Technology (MEXT), High Energy Accelerator Research
Organization (KEK) and Japan Aerospace Exploration Agency (JAXA) in Japan, and
the K.~A.~Wallenberg Foundation, the Swedish Research Council and the
Swedish National Space Board in Sweden.
 
Additional support for science analysis during the operations phase is gratefully
acknowledged from the Istituto Nazionale di Astrofisica in Italy and the Centre
National d'\'Etudes Spatiales in France. This work performed in part under DOE
Contract DE-AC02-76SF00515.

This work is part of a project that has received funding from the European Union's Horizon 2020 Research and Innovation Programme under the Marie Sk\l{}odowska-Curie grant agreement NO 664931.
\end{acknowledgements}


\bibliographystyle{aa}
\bibliography{OptChar} 

\begin{appendix} 
\section{New BZB candidates}
\label{sec.foo}

\begin{figure*}
    \centering
    \includegraphics[width=\linewidth]{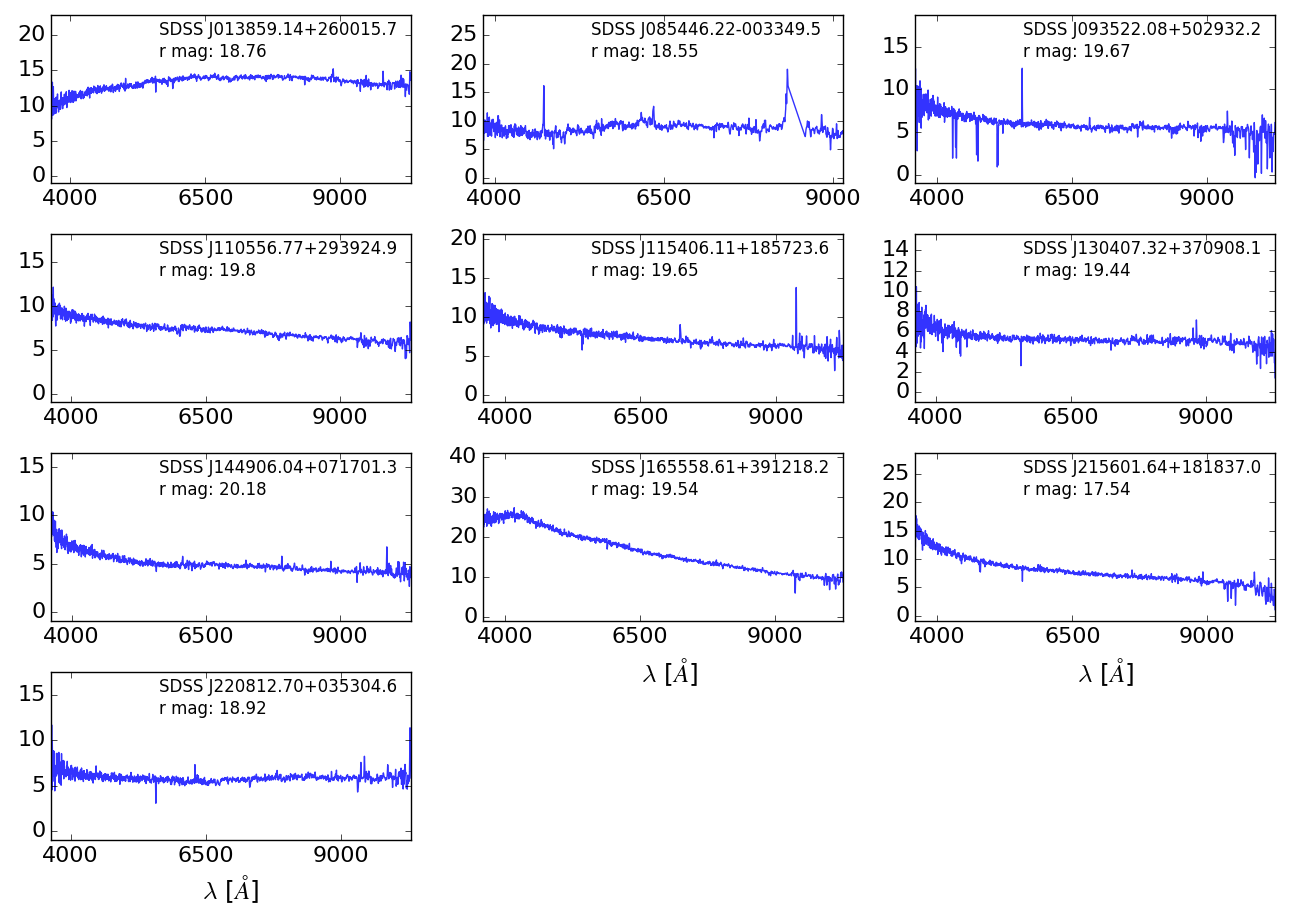}
    \caption{SDSS smoothed optical spectra of the new BZBs in WIBRaLS2. All sources present featureless spectra with color index $u-r < 1.4$. We also show their SDSS name and r magnitude.}
    \label{fig:NewBllacs_1}
\end{figure*}

\begin{figure*}
    \centering
    \includegraphics[width=\linewidth]{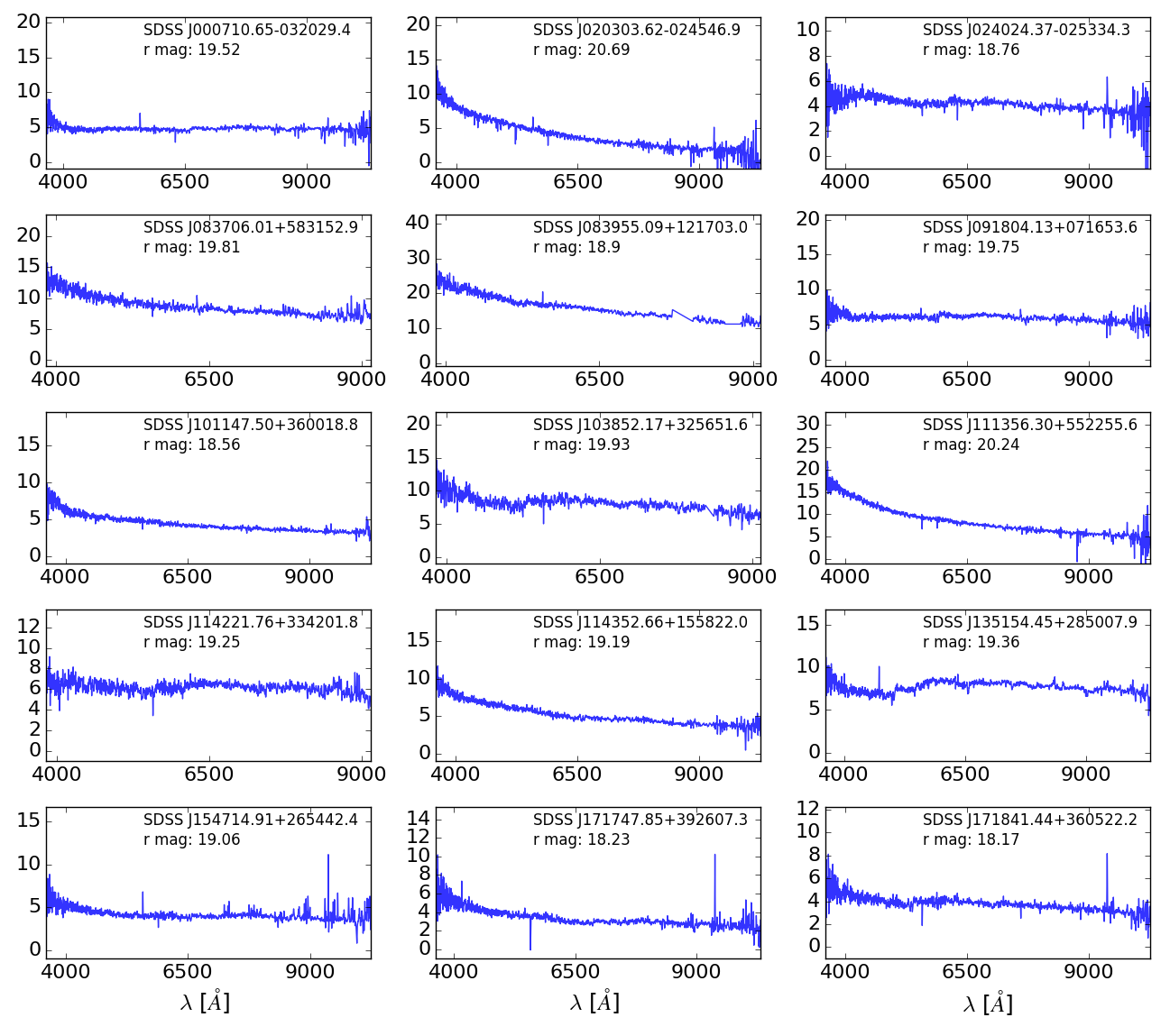}
    \caption{SDSS smoothed optical spectra of the new BZBs in KDEBLLACS. All sources present featureless spectra with color index $u-r < 1.4$. We also show their SDSS name and r magnitude.}
    \label{fig:NewBllacs_2}
\end{figure*}

During the spectral classification process described in Section \ref{SpecClass}, we found 25 new BZBs not available in Roma-BZCAT or during our follow up spectroscopic campaign \citep[see e.g.][]{paggi2014optical,landoni2015optical,ricci2015optical,crespo2016opticalV,crespo2016opticalVI,massaro2016BlazarQuest,pena2017optical,marchesini2019optical,pena2019optical}. Their smoothed optical SDSS DR15 spectra (SNR > 5) are available in Figures \ref{fig:NewBllacs_1} and \ref{fig:NewBllacs_2} as well as their r band magnitudes.

\end{appendix}

\end{document}